\documentclass[twocolumn,showpacs,aps,preprintnumbers,amsmath,amssymb,superscriptaddress]{revtex4}
\usepackage{amsmath}
\usepackage{graphicx}
\usepackage{subfig}
\usepackage{amssymb}
\usepackage{bm}
\usepackage{color}
\usepackage{makecell}
\usepackage[justification=RaggedRight]{caption}

\begin{document}
\pagestyle{plain}
\title{Existence problem of proton semi-bubble structure in the $2_1^+$ state of $^{34}$Si}
\author{Feng Wu}
\affiliation{China Institute of Atomic Energy, Beijing 102413, China}
\affiliation{Key Laboratory of Radiation Physics and Technology of Ministry of Education,
	School of Physics Science and Technology, Sichuan University, Chengdu 610065, China}
\author{C.L. Bai}
\email[]{bclphy@scu.edu.cn}
\affiliation{Key Laboratory of Radiation Physics and Technology of Ministry of Education,
	School of Physics Science and Technology, Sichuan University, Chengdu 610065, China}
\author{J.M. Yao}
\affiliation{Department of Physics and Astronomy, University of North Carolina, Chapel Hill, North Carolina 27516-3255, USA}
\affiliation{School of Physical Science and Technology, Southwest University, Chongqing 400715, China\\}
\author{H.Q. Zhang}
\email[]{huan@ciae.ac.cn}
\affiliation{China Institute of Atomic Energy, Beijing 102413, China}
\author{X.Z. Zhang}
\affiliation{China Institute of Atomic Energy, Beijing 102413, China}

\begin{abstract}
The fully self-consistent Hartree-Fock (HF) plus random phase 
approximation (RPA) based on Skyrme-type interaction is used to study the existence problem of proton semi-bubble structure in the $2_1^+$ state of $^{34}$Si. 
The experimental excitation energy and the B(E2) strength of the $2_1^+$ state in $^{34}$Si can be reproduced quite well.  The tensor effect is also studied. 
It is shown that the tensor interaction has a notable impact on the excitation energy of the $2_1^+$ state and a small effect on the B(E2) value. Besides, its
 effect on the density distributions in the ground and $2_1^+$ state of $^{34}$Si is negligible. Our present results with T36 and T44 show that the $2_1^+$ state of $^{34}$Si is mainly caused by proton transiton from $\pi 1d_{5/2}$ orbit to $\pi 2s_{1/2}$ orbit, and the existence of a proton semi-bubble structure in this state is very unlikely.
\end{abstract}


\maketitle
\thispagestyle{plain}
\section{Introduction} \label{introduction}
The discussion of new topology structure of nuclei can date back to H. A. Wilson 
\cite{Wilson46} and J. A. Wheeler \cite{Wheeler} prior to 1950s. The bubble 
structure is one of the simplest but widely discussed topology structure in history \cite{Wong72,Davies73,Campi73,Beiner73,Saunier74,Nilsson74}, 
and it is still a hot topic both theoretically and experimentally in recent 
years \cite{Khan08,Grasso09,Chu10,Yao12,Yao13,XYWu14,Nakada13,Mutschler17,Duguet17,Decharge99,Bender99,Nazarewicz02,Decharge03,Wang11,Wang13,Wang15}. However, 
a bubble structure is not expected to exist in nuclei because 
the nature of nucleon-nucleon interaction will lead to a saturation density ($ 
\rho_{sat} \sim 0.16$ fm$^{-3}$) at the center of nucleus. Although the ``true-bubbl'' structure with vanishing density at the center does not exist in realistic atomic nuclei, there is a possibility that a ``semi-bubble'' structure with a significantly reduced density in nuclear interior can be formed in some nuclei with particular configurations.  This is also the main subject of this work.

The occurrence of a (semi-)bubble in a spherical nucleus is a direct result of the 
nucleon deficiency in the $s_{1/2}$ orbital, which is the only one that 
contributes to the central density. Theoretically, a (semi-)bubble structure 
can exist in many nuclei, ranging from intermediate-mass isotopes to hyperheavy 
systems, including $^{34}$Si \cite{Grasso09,Nakada13,Duguet17}, $^{44}$S \cite{Chu10}, $^{46}$Ar \cite{Khan08,Chu10}, neutron-rich Ar isotopes around $^{68}$Ar \cite{Khan08}, $^{36}$Ar and Hg isotopes \cite{Campi73,Davies73,Wong72}, superheavy and hyperheavy isotopes \cite{Bender99,Nazarewicz02,Decharge99,Decharge03}, etc. Some of these candidates, such 
as $^{36}$Ar and Hg isotopes, have been ruled out by experiments \cite{Yao12}. However, other candidates are still standing. 
Especially, the proton semi-bubble structure in $^{34}$Si has been consistently predicted in many works by using different models, 
such as shell model (SM), nonrelativistic and relativistic microscopic mean-field approaches \cite{Grasso09, Nakada13}, and ab initio self-consistent
Green's function many-body method \cite{Duguet17}. Experimentally, A. Mutschler et al. \cite{Mutschler17} recently found that the $2s_{1/2}$ 
proton orbit in $^{34}$Si is essentially empty by using the one-proton removal 
reaction technique. Their work provides the first 
experimental evidence for the existence of a proton semi-bubble structure in 
the ground state of $^{34}$Si and also motivates us to study the same problem in its 
excited state.  
 
Most of the works mentioned above are concentrated on the ground state of the
candidate nuclei. For the low-lying states some researches were reported in Refs. \cite{Wong72,Yao13,XYWu14}. In these works, 
different methods were applied, such as the Strutinski's theory \cite{Wong72} and the projected generator coordinate model (GCM) \cite{Yao13,XYWu14}. 
In principle, models 
such as time-dependent Hartree-Fock (TDHF) and RPA maybe also valid in this topic. But so far, there is no report about the research
on this topic done by these models. In this work, the fully 
self-consistent HF+RPA model with Skyrme interactions is adopted to study the existence 
problem of the proton semi-bubble structure in the low-lying state of 
the ``doubly-magic'' nucleus $^{34}$Si \cite{Sorlin08}. 

It is well-known that RPA is the small amplitude limit of the TDHF. So, in principle, the spherical HF+RPA method can only be applied to describe excited state 
with small deformation. Although the big deformed 0$_2^+$ state of $^{34}$Si has been detected in experiment \cite{Rotaru12}, it is now beyond our reach to reproduce this state. 
Thus, in this work, we will limit ourselves to the $2_1^+$ state of $^{34}$Si, which might be well reproduced in the HF+RPA model.

Moreover, we are also interested in the role played by the tensor interaction on the bubble 
structure in the low-lying state of $^{34}$Si, as the tensor interaction was reported to have
strong effect on the low-lying collective state of nuclei \cite{Bai09,Bai09b,Bai14,Cao09,Cao11}.
In addition, the effect of the Skyrme tensor force on the bubble stucture in ground state was also
studied through the HF or Hartree-Fock-Bogliubov (HFB) method in Refs. \cite{Wang11,Wang13,Wang15}.
We deem interesting to study if the tensor interaction has an effect on the proton semi-bubble structure in
the low-lying state of $^{34}$Si.

This paper is organized as follows. In Sec. \ref{formalism}, 
a few necessary details about
the method is presented. Sec. \ref{numdetail} displays the numerical details in the present calculations. The results and discussion are presented in Sec. 
\ref{result}. Summary and prospect are made in Sec. \ref{summary}.

\section{Formalism}\label{formalism}
The basic ideas and details of HF and RPA can be found in a mass of textbooks 
\cite{Ring80,Rowe10} and papers \cite{Vautherin72, Colo13}. Here, we will only 
give some formulas that are necessary for explanation.

In spherical case, the wave function of a single particle (s.p.) state (denoted 
by $\alpha$) with principal quantum number $n_{\alpha}$, orbital and total 
angular momentum $l_{\alpha},j_{\alpha}$, total angular momentum projection 
$m_{\alpha}$, and isospin $q_{\alpha}$ can be written as:
\begin{equation}\label{spwf}
\phi_{\alpha}^{q_{\alpha}}\left( \vec{r},\sigma\right)=R_{\alpha}\left(r\right)\left[Y_{l_{\alpha}}\left(\hat{r}\right) \otimes 
\chi_{1/2}\left(\sigma\right)\right]_{j_{\alpha}m_{\alpha}}\chi_{q_{\alpha}}\left(\tau\right)
\end{equation}
where $\chi_{1/2}\left(\sigma\right)$ and $\chi_{q_{\alpha}}\left(\tau\right)$ 
denote the two-component spinor in the spin and isospin space, respectively.  

The creation and annihilation operators of the RPA phonons with angular momentum $JM$ are defined as:
\begin{eqnarray} \label{phoperator}
&&A^{\dagger}_{mi}\left( JM\right)=\sum\limits_{m_{m} m_{i}} \left\langle j_{m} m_{m} j_{i} m_{i} \left|\right. JM \right\rangle a_{j_m m_m}^{\dagger} a_{j_i \overline{m_i}} \nonumber \\
&&A_{mi}\left( J \overline{M}\right)=(-1)^{J+M} A_{mi}\left( J{-M}\right)
\end{eqnarray}
where $a_{\alpha}^{\dagger}$ and $a_{\alpha}$ are the creation and annihilation operator
of $\alpha$ s.p. state. And in this paper, we will always use $m, n$ to denote 
unoccupied states and $i, j$ to denote occupied states.

The excitation operators in RPA are linear combination of the particle-hole (ph)
creation and annihilation operators:
\begin{eqnarray} \label{phononoperator}
O_{\lambda JM}^{\dagger} =\sum\limits_{mi} \left\{ X_{mi}^{\lambda} \left( J \right) A_{m i}^{\dagger} \left(JM\right) -Y_{mi}^{\lambda}\left( J \right) {A}_{mi}\left( J\overline{M}\right)  \right\}
\end{eqnarray}
where $\lambda JM$ denotes different excited states with total angular momentum $JM$, $X$ and $Y$ denote the forward and backward amplitudes, respectively. And the RPA matrix equation is:
\begin{eqnarray} \label{RPAeq}
\begin{pmatrix} A&B\\B^{*}&A^{*}\end{pmatrix}\begin{pmatrix}X^{\lambda}\\Y^{\lambda}\end{pmatrix}=E_{\lambda}\begin{pmatrix} 1&0\\0&-1\end{pmatrix}\begin{pmatrix}X^{\lambda}\\Y^{\lambda}\end{pmatrix}
\end{eqnarray}
where $E_{\lambda}$ is the excitation energy of the $\lambda$th excited state. And $A$, $B$ are the matrix elements derived from the second derivatives of the Skyrme energy functional \cite{Colo13}.

The density of excited state $\left| \lambda JM \right\rangle$ in RPA model is \cite{Terasaki}:
\begin{eqnarray} \label{densityf}
\rho_{\lambda}^q \left( \vec{r}\right) &= &\left\langle \lambda JM\left| \hat{\rho}_q \left(\vec{r}\right) \right| \lambda JM \right\rangle \nonumber\\
&=&\left\langle RPA \left|O_{\lambda JM} \hat{\rho}_q O_{\lambda JM}^{\dagger} \right| RPA \right\rangle \nonumber\\
&\approx &\left\langle HF \left| \left[O_{\lambda JM}, \left[ \hat{\rho}_q \left( \vec{r} \right) , O_{\lambda JM}^{\dagger} \right] \right] \right| HF \right\rangle \nonumber \\
&&+\left\langle HF \left|\left[O_{\lambda JM}, O_{\lambda JM}^{\dagger} \right] \hat{\rho}_q \left( \vec{r}\right) \right| HF \right\rangle
\end{eqnarray}
where $\left|HF\right\rangle$ and $\left| RPA\right\rangle$ represent the HF 
vacuum and RPA vacuum, respectively, and $q$ denotes either proton or neutron. 
To obtain equation (\ref{densityf}), we have assumed the so-called ``quasiboson 
approximation". The density distribution of excited state $\left|\lambda 
JM\right\rangle$ can be easily derived by expanding formula (\ref{densityf}) and 
using the s.p. wave function in formula (\ref{spwf}). And it can be expressed as a 
sum of multipole components:  
\begin{eqnarray} \label{densitycom}
\rho_{\lambda}^q \left( \vec{r}\right) &=& \rho_{HF}^q \left( r\right) + \Delta \rho_{\lambda}^q \left( \vec{r}\right) \nonumber\\
&\approx &\rho_{HF}^q \left( r\right) + \sum\limits_{ \begin{tiny} \begin{array}{l} \;\;\;\;L \geq 0 \\\;L\;is\;even \end{array} \end{tiny}} \rho_{\lambda L}^q \left(r,cos \theta\right) \nonumber\\
&\approx& \rho_{HF}^q \left( r\right) + \sum\limits_{ \begin{tiny} 
\begin{array}{l} \;0 \leq L \leq 2J \\\;\;L\;is\;even \end{array} 
\end{tiny}} \rho_{\lambda L}^q \left(r,cos \theta\right)
\end{eqnarray}
where $\rho_{HF}^q \left( r \right) $ is the ground state density distribution, 
and $\rho_{\lambda L}^q \left(r,cos \theta\right) \propto Y_{L0} \left(\hat{r}\right)$. When 
the ``quasiboson approximation" is accurate enough, the $L>2J$ components should 
be very small, which is the reason for the last equation in formula 
(\ref{densitycom}). In fact, the $L>2J$ components in the present calculations are smaller than $10^{-11}$ fm$^{-3}$.

Formula (\ref{densitycom}) suggests that the density distribution of the 
$J>0$ excited states are actually axially symmetric in the HF+RPA 
framework. Our calculations show that the density distribution of the $2_1^+$ 
state of $^{34}$Si at the center area is almost spherically symmetric. Therefore, it is convenient
to define an average density to study the semi-bubble structure:
\begin{eqnarray}\label{densityave}
\bar{\rho}_{\lambda}^q (r)&=&\dfrac{\int \rho_{\lambda}^q \left( \vec{r}\right) d \Omega }{4 \pi} \nonumber\\
&=& \sum\limits_{mni}\dfrac{\delta_{j_m j_n} \delta_{l_m l_n}}{4\pi}\left(X_{mi}^{\lambda}X_{ni}^{\lambda}+Y_{mi}^{\lambda}Y_{ni}^{\lambda}\right) R_m(r)R_n(r)\nonumber\\
&&-\sum\limits_{mij}\dfrac{\delta_{j_i j_j} \delta_{l_i l_j}}{4\pi}\left(X_{mi}^{\lambda}X_{mj}^{\lambda}+Y_{mi}^{\lambda}Y_{mj}^{\lambda}\right) R_i(r)R_j(r)\nonumber\\
&&+\rho_{HF}^q \left( r\right)
\end{eqnarray}
in which the sum runs through either proton or neutron states. 

We are also interested in effect of the tensor interaction. And the zero-range Skyrme tensor term reads \cite{Skyrme59,Bai09}:
\begin{eqnarray} \label{tensorf}
V_{T}&=&\dfrac{T}{2}\{[(\mathbf{\sigma _{1}\cdot {k}^{\prime
	})(\sigma _{2}\cdot {k}^{\prime })-\dfrac{1}{3}\left( \sigma
	_{1}\cdot \sigma _{2}\right) {k}^{\prime 2}]\delta(r) }\nonumber\\
&&+\delta (\bf {r})[ (\bf{\sigma _{1}\cdot {k})(\sigma _{2}\cdot
	{k})-\dfrac{1}{3}\left(
	\sigma _{1}\cdot \sigma _{2}\right) {k}^{2}}] \}\nonumber\\
&&+\dfrac{U}{2}\{\left( \sigma _{1}\cdot \bf{k}^{\prime }\right)
\delta (\bf{r}) (\sigma _{2}\cdot \bf{k}) +\left( \sigma _{2}\cdot
\bf{k}^{\prime }\right) \delta (\bf{r})(\sigma _{1}\cdot
{k})\nonumber\\
&&-\dfrac{2}{3}\left[ (\bf{\sigma }_{1}\cdot \bf{\sigma }_{2})
\bf{k}^{\prime }\cdot \delta (\bf{r})\bf{k} \right] \}.\label{tensor}
\end{eqnarray}
where $T$ and $U$ are the strengths of triplet-even and triplet-odd tensor terms, respectively, $\bf{r}=\bf{r_1}-\bf{r_2}$, and the operator $\mathbf{k=(\nabla_1 -\nabla_2)/2i}$ acts on the right and $\mathbf{k^{\prime}=-(\nabla_1^{\prime} -\nabla_2^{\prime})/2i}$ acts on the left.

To compare the reduced transition strength $B(E2;0_1^+\rightarrow2_1^+)$ with the experimental result, the transition operator for $J^{\pi}=2^{+}$ we use is:
\begin{eqnarray} \label{transoperator}
\hat{F}_{2M}&=&\sum\limits_{i=1}^{Z} e r_i^2 Y_{2M} \left( \Omega_i \right)
\end{eqnarray}
in which the sum only runs through protons.

The reduced transition strength can be calculated by
\begin{eqnarray}
B(EJ;0_1^+\rightarrow\lambda JM)=\left|\sum\limits_{mi}\left(X_{mi}^{\lambda}+Y_{mi}^{\lambda}\right)\left\langle m \left|\left|\hat{F}_J\right|\right| i\right\rangle\right|^2
\end{eqnarray}

To evaluate the contribution of a particular ph configuration $nj$ to the reduced transition strength, we define a weight factor for this configuration:
\begin{eqnarray}
W_{nj}^{\lambda}=\frac{\left(X_{nj}^{\lambda}+Y_{nj}^{\lambda}\right)\left\langle n \left|\left|\hat{F}_J\right|\right|j\right\rangle}{\sum\limits_{mi}\left(X_{mi}^{\lambda}+Y_{mi}^{\lambda}\right)\left\langle m \left|\left|\hat{F}_J\right|\right| i\right\rangle}
\end{eqnarray}
The sum of this factor for all configurations is equal to 1, so it can appropriately reflect the contribution of each configuration.

\section{NUMBERICAL DETAILS}\label{numdetail}
The calculations are started by solving the HF equations in coordinate space with a radial mesh extending up to 20 fm in a step of 0.1 fm. And a cutoff of s.p. energy by 80 MeV is adopted in RPA calculation.

In the present calculations, the Skyrme type interactions T36 and T44 \cite{Lesinski07} are employed, as recommended in Ref. \cite{Cao11}. The tensor terms are manually switched on (denoted as w/) and off (denoted as w/o) simultaneously in HF and RPA. Other terms of the interaction, including the central term, the two-body spin-orbit interaction, and the Coulomb interaction, are included in both HF and RPA in all calculations. So the present calculations are fully self-consistent. And the energy weighted sum rule (EWSR) in the present calculations agrees quite well (the deviation is less than 0.2\%) with the analytical ones. 

\section{Results and discussion}\label{result}
\begin{figure}[!hbt]
	\centering
	\includegraphics[height = 6.0cm, width = 8.0 cm]{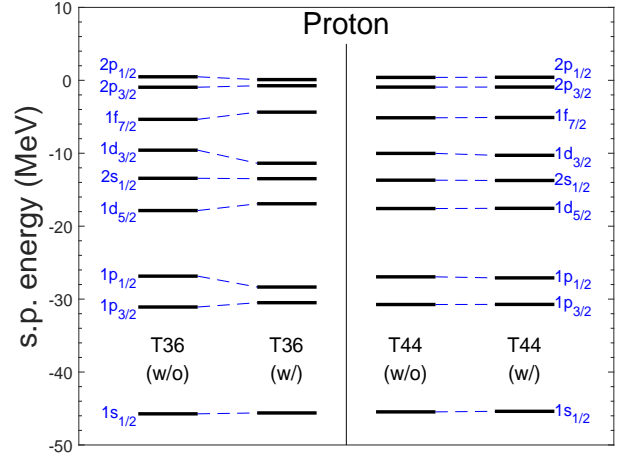}
	\caption{Proton s.p. energy levels calculated by employing T36 and T44. The tensor terms are manually switched on and off. The label (w/o) and (w/) denotes the results calculated without and with tensor interaction, respectively.}
	\vspace{-1em}
	\label{spE}
\end{figure}
The proton s.p. energy levels of $^{34}$Si calculated with HF in both T36 and T44 cases are displayed in Fig. \ref{spE}. It is shown that the tensor interaction has a significant impact on the s.p. levels by influencing the spin-orbit splitting, as been pointed out in Ref. \cite{Colo07,Brink07}. It is also obvious to see that the tensor terms in T36 has much more remarkable influence on the s.p. energy in $^{34}$Si than that in T44. In T36, the energy splitting of the proton $1p_{1/2}$ and $1p_{3/2}$ states is reduced from 4.23 MeV to 2.15 MeV by the tensor terms while in T44 the value is reduced from 3.80 MeV to 3.64 MeV. We note that the reduction of the spin-orbit splitting in T36(w/) case is mostly contributed from the tensor force, instead of the change in the spin-orbit field which is related to the density distribution. We will show later that the tensor terms turn out to have a marginal effect on the density distribution of both ground state and $2_1^+$ state. The different levels of tensor effect in T36 and T44 can be easily understood by comparing the triplet-even and triplet-odd strengths for the tensor interaction, i.e. $T$ and $U$ in formula (\ref{tensorf}), in these two parameter sets. In T36, $[T,U]$=[27.2,341.8], while in T44, the values are [521.0,21.5]. Since $^{34}$Si is spin-orbit saturated for neutron, according to formula (4) in Ref. \cite{Colo07}, the triplet-odd term will have much more important influence on the proton spin-orbit potential than the triplet-even term. It is also shown that the order of the s.p. levels keeps unchanged with both T36 and T44. Therefore, the density distribution in the ground state will not be significantly influenced by the tensor terms, and the bubble property of the ground state will also keep unchanged.

The excitation energy and the reduced transition strength B(E2;$0_1^+ \rightarrow 2_1^+$) in both T36 and T44 cases are dispalyed in Table \ref{tableEB}. It can be noted that when the tensor terms are included, both T36 and T44 can well reproduce the experimental energy and the B(E2) value. It is also shown that the tensor interaction has a significant
\begin{table}[!htbp]
	\vspace{-1em}
	\centering 
	\caption{Energy in units of MeV, B(E2;$0_1^+ \rightarrow 2_1^+$) in units of 
		$e^2$ fm$^4$ with T36 and T44. The tensor terms are manually switched on and off. The experimental results are taken from \cite{Nummela01,Ibbotson98}.}
	\setlength{\tabcolsep}{4pt}
	\renewcommand{\arraystretch}{1.2}
	\begin{tabular}{lccccc}
		\toprule
		\;&T36(w/o) &T36(w/)&T44(w/o)&T44(w/)&exp\\
		\hline
		$E_{2_1^+}$ & 4.142 & 3.299 & 3.643 & 3.429 &3.326 \\
		B(E2)  & 109.5 & 104.0 & 104.8 & 99.3  &85(33) \\
		\botrule
	\end{tabular}
	\label{tableEB}
\end{table}
\begin{table}[!htbp]
	\caption{The diagonal ph matrix elements A(B) in Eq. (\ref{RPAeq}) of the most important configuration $\pi 2s_{1/2}$ $\otimes$ $\pi 1d_{5/2}$ for the $2_1^+$ state of $^{34}$Si. $M_{central}$, $M_{tensor}$, $M_{so}$, $M_{Coul}$ denote the contribution of the central part, the tensor parts, the two body spin-orbit part, and the Coulomb part, respectively.}
	\setlength{\tabcolsep}{3pt}
	\renewcommand{\arraystretch}{1.2}
	\begin{tabular}{lcccc}
		\toprule
		\;&$M_{central}$ &$M_{tensor}$ &$M_{so}$& $M_{Coul}$\\
		\hline
		T36& 0.590(0.547) &0.217(0.000)&0.198(0.000)&0.020(0.020) \\
		T44& 0.668(0.690) &0.014(0.000)&0.197(0.000)&0.020(0.020)\\
		\botrule
	\end{tabular}
	\label{tableME}
\end{table}
\noindent impact on the excitation energy of the $2_1^+$ state while its effect on the B(E2) value is not strong, which is similar to the results of $^{48}$Ca and $^{208}$Pb in Ref. \cite{Cao09}.

The different parts of the diagonal RPA matrix elements for the most important configuration $\pi 2s_{1/2}$ $\otimes$ $\pi 1d_{5/2}$ are displayed in Table \ref{tableME}. It is shown that the tensor effect in T36 is much more notable than that in T44 in the RPA calculation for the $2_1^+$ state.  However, in both cases,  
\begin{figure}[!hbt]
	\vspace{1em}
	\centering
	\includegraphics[height = 6.0cm, width = 8.0 cm]{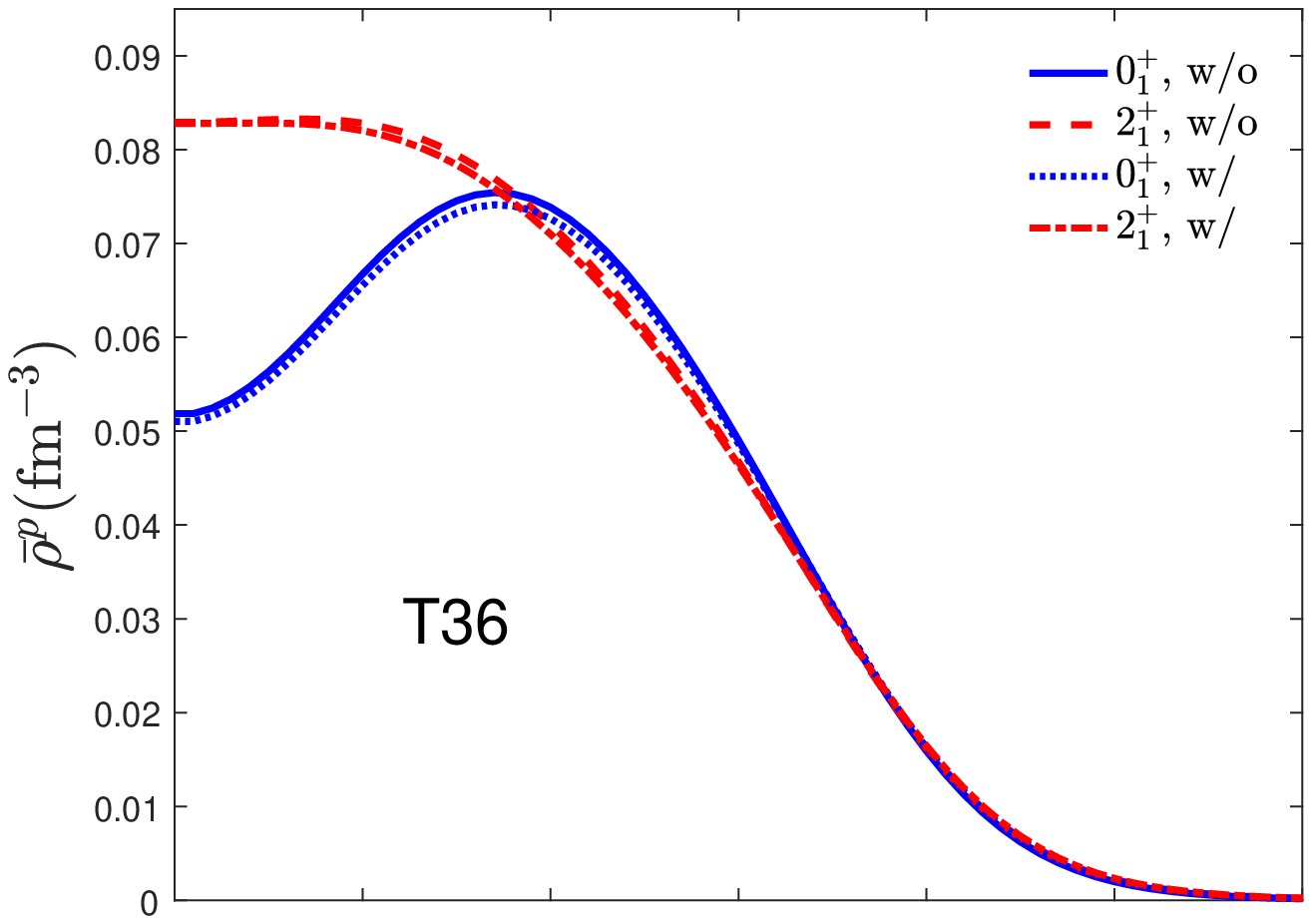}\\
	\vspace{1.3em}
	\includegraphics[height = 6.0cm, width = 8.0 cm]{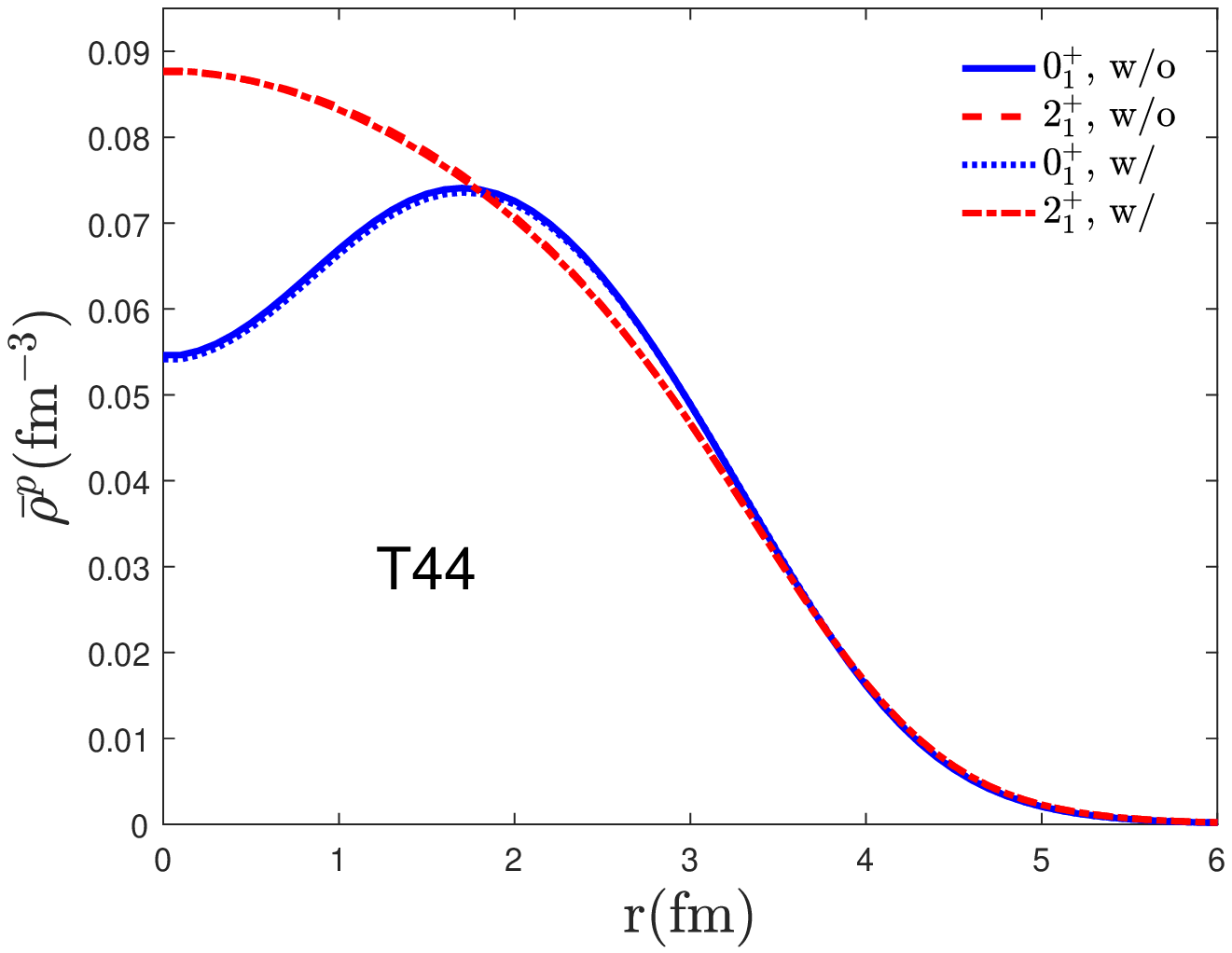}
	\caption{The average proton density distributions in ground ($0_1^+$) state and $2_1^+$ state with T36 and T44. The tensor terms are manually switched on and off.}
	\vspace{-1em}
	\label{rhop}
\end{figure} 
\begin{table}[!htbp]
	\caption{Amplitude of the $\pi 2s_{1/2}$ $\otimes$ $\pi 1d_{5/2}$ 
		configuration, the weight factor for B(E2) of this configuration (denoted as $W_{nj}^{\lambda}$), and depletion factors for the ground ($0_1^+$) and 
		$2_1^+$ state in $^{34}$Si with T36 and T44. The tensor terms are manually switched on and off.}
	\setlength{\tabcolsep}{5pt}
	\renewcommand{\arraystretch}{1.5}
	\begin{tabular}{lccccc}
		\toprule
		\;&T36(w/o) &T36(w/)&T44(w/o)&T44(w/)\\
		\hline
		$X^2-Y^2$          & 0.965 &0.976&0.967&0.963\\
		$W_{nj}^{\lambda}$ & 0.746 &0.766&0.757&0.757\\
		$F_{max}^{0_1^+}$  & 31.2\% &31.2\%&26.2\%&26.4\%\\
		$F_{max}^{2_1^+}$  & 0.4\% &0.5\%&0&0 \\
		\botrule
	\end{tabular}
	\vspace{-1em}
	\label{tableXY}
\end{table}
\noindent the tensor matrix elements are relatively small compared with those of the central term. Therefore, the tensor effect in the RPA calculation is relatively small. The two body spin-orbit, and the Coulomb matrix elements are also displayed in Table \ref{tableME} for reference. They are small compared with those of the central term.
 
Fig. \ref{rhop} displays the mean proton density distribution with T36 and T44. For the ground state, the results of T36 and T44 both indicate a notable proton semi-bubble structure. It can be noted that the tensor interaction has a negligible effect on the density distribution in both cases, since it does not change the order of the s.p. levels and only contributes to the spin-orbit potential in HF. For the $2_1^+$ state, the present results indicate that the proton semi-bubble structure is very unlikely, which is consistent with the previous study with GCM method based on a relativistic energy functional \cite{Yao13}. Moreover, the tensor effect on the density distribution in the $2_1^+$ state is also negligible, as the contribution of the tensor interaction to the RPA matrix elements is relatively small.

To describe the depletion quantificationally, we use a depletion factor \cite{Yao13}:
\begin{equation} \label{depletionf}
F_{max}=\dfrac{\rho_{max,p} -\rho_{cent,p}}{\rho_{max,p}}
\end{equation}
with $\rho_{cent,p}$ and $\rho_{max,p}$ being the center density and maximal density, respectively.

The amplitude of the most important configuration $\pi 2s_{1/2}$ $\otimes$ $\pi 1d_{5/2}$, the weight factor for B(E2) of this configuration and the depletion factors of the ground state and $2_1^+$ state are listed in Table \ref{tableXY}. It can be seen that the tensor interaction does not significantly influence the amplitude of this important configuration. As a result, it does not significantly influence the bubble property in the $2_1^+$ state, which is reflected on the depletion factor. It is also shown that this important configuration contributes most to the B(E2) strength. Therefore, we are much more confident to draw the conclusion that the $2_1^+$ state of $^{34}$Si is mainly caused by proton transiton from $\pi 1d_{5/2}$ orbit to $\pi 2s_{1/2}$ orbit. A large amplitude of this configuration indicates that the $\pi 2s_{1/2}$ orbit has a large possibility to be occupied. Consequently, the proton semi-bubble is filled and disappears in the $2_1^+$ state.
 
\section{SUMMARY AND PROSPECTS}\label{summary}
The fully self-consistent HF+RPA method has been used to study the proton semi-bubble structure in the $2_1^+$ state of $^{34}$Si. The present results with T36 and T44 can well reproduce the experimental energy and B(E2) value. A very large amplitude of $\pi 2s_{1/2}$ $\otimes$ $\pi 1d_{5/2}$ configuration indicates that the $2_1^+$ state of $^{34}$Si is mainly caused by proton transiton from $\pi 1d_{5/2}$ orbit to $\pi 2s_{1/2}$ orbit. As a result, the proton semi-bubble structure is filled and is unlikely to exist in the $2_1^+$ state of $^{34}$Si.

The tensor effect on the proton semi-bubble structure in the $2_1^+$ state of $^{34}$Si has been studied by manually removing the tensor terms from T36 and T44. It is shown that the tensor interaction has a significant effect on the s.p. levels and the excitation energy of the $2_1^+$ state in $^{34}$Si. But its effects on the B(E2) value and the density distribution of the ground and $2_1^+$ state are very small.

As mentioned in the introduction part, a (semi-)bubble structure is expected to 
exist in many nuclei, including some heavy, superheavy and hyperheavy systems. 
To extend the study to these nuclei, a projected Hartree-Fock-Bogolyubov 
(PHFB) plus projected quasiparticle random phase approximation (PQRPA) method is 
needed to develop in the future. Systematic calculations are also needed to 
search for (semi-)bubble structure or other exotic density distribution in 
excited states of different nuclei.
 
\begin{acknowledgments}
This work is supported by the National Natural Science Foundation of China under Grant Nos. 11375266, 11575120, and 11575148.
\end{acknowledgments}

\end{document}